\documentclass[onecolumn]{article}

\usepackage{lineno}
\usepackage{nopageno}
\usepackage{authblk}
\usepackage{graphicx}
\usepackage[explicit]{titlesec}
\usepackage[labelfont=bf,labelsep=endash,font=footnotesize]{caption}
\usepackage{tabu}
\usepackage{xcolor}
\usepackage[hidelinks]{hyperref}
\usepackage[hang]{footmisc}
\usepackage[normalem]{ulem}
\usepackage[top=2.5cm, bottom=2.8cm, left=1.5cm, right=1.5cm]{geometry}
\usepackage{abstract}
\usepackage{mathtools}
\usepackage{amsmath}
\usepackage{amssymb}
\usepackage{balance}

\usepackage{multirow}
\usepackage{pgfplots}
\usepackage{tikz}
\usepackage{standalone}

\usepackage{algorithm}
\usepackage{algpseudocode}

\DeclareMathOperator*{\argmax}{arg\,max}
\DeclareMathOperator*{\argmin}{arg\,min}
\renewcommand{\vec}[1]{\boldsymbol{#1}}
\newcommand{\norm}[1]{\left|\left|#1\right|\right|}

\def\layersep{2cm}
\def\layersep{2.75cm}

\makeatletter
\renewcommand\AB@affilsepx{, \protect\Affilfont}
\makeatother

\providecommand{\keywords}[1]{\textbf{Keywords}\ \ \textendash\ \   #1}

\titleformat{\section}{\large\bfseries}{\thesection.}{1em}{\MakeUppercase{#1}}
\titlespacing*{\section}{0pt}{12pt}{6pt}

\titleformat{\subsection}{\large}{\thesubsection}{1em}{#1}
\titlespacing*{\subsection}{0pt}{12pt}{6pt}

\titleformat{\subsubsection}{\large\itshape}{\thesubsubsection}{1em}{#1}
\titlespacing*{\subsubsection}{0pt}{12pt}{6pt}

\newcommand{\ITUurl}[1]{\textcolor{blue}{\urlstyle{same}\url{#1}}}

\newcommand{\ITUpar}{\vspace{8pt}\par}

\renewenvironment{abstract}
               {\list{}{
               \setlength{\rightmargin}{0mm}
               \setlength{\leftmargin}{0mm}
               \vspace{-0.25in}
                \item[\textit{\textbf{\hspace{22pt}Abstract  }}  \textendash]\relax}}
               {\endlist}
\def\starttable{\vspace{6pt}\begin{table}[ht]\center}
\def\startfigure{\vspace{6pt}\begin{figure}[ht]\center}

\makeatletter
\def\tagform@#1{\maketag@@@{\ignorespaces#1\unskip\@@italiccorr}}
\makeatother

\usepackage{caption}
\usepackage{subcaption}

\title{\large{\textbf{\uppercase{Federated Spatial Reuse Optimization in Next-Generation Decentralized IEEE 802.11 WLANs}}}}

\author[1]{\normalsize{Francesc~Wilhelmi}}
\author[2]{\normalsize{Jernej~Hribar}}
\author[3]{\normalsize{Selim~F.~Yilmaz}}
\author[3]{\normalsize{Emre~Ozfatura}}
\author[3]{\normalsize{Kerem~Ozfatura}}
\author[4]{\normalsize{Ozlem~Yildiz}}
\author[3,5]{\normalsize{Deniz~G\"{u}nd\"{u}z}}
\author[6]{\normalsize{Hao~Chen}}  
\author[6]{\normalsize{Xiaoying~Ye}}  
\author[6]{\normalsize{Lizhao~You}}
\author[3]{\normalsize{Yulin~Shao}}
\author[1]{\normalsize{Paolo~Dini}}
\author[7]{\normalsize{Boris~Bellalta}}

\affil[1]{\normalsize{CTTC (Spain)}}
\affil[2]{\normalsize{CONNECT~Centre, Trinity~College~Dublin (Ireland)}}
\affil[3]{\normalsize{Imperial~College~London (United Kingdom)}}
\affil[4]{\normalsize{New~York~University (USA)}}
\affil[5]{\normalsize{University of Modena and Reggio Emilia (Italy)}}
\affil[6]{\normalsize{Xiamen~University (China)}}
\affil[7]{\normalsize{Universitat~Pompeu~Fabra (Spain)}}

\date{\vspace{-12pt}{\small NOTE: Corresponding author: Francesc Wilhelmi, francesc.wilhelmi@cttc.cat} \\  \endgraf\rule{\textwidth}{1pt}}

\begin{document}
	
	
	
	\maketitle
	
	\begin{abstract}
		\textit{As wireless standards evolve, more complex functionalities are introduced to address the increasing requirements in terms of throughput, latency, security, and efficiency. To unleash the potential of such new features, artificial intelligence (AI) and machine learning (ML) are currently being exploited for deriving models and protocols from data, rather than by hand-programming. In this paper, we explore the feasibility of applying ML in next-generation wireless local area networks (WLANs). More specifically, we focus on the IEEE 802.11ax spatial reuse (SR) problem and predict its performance through federated learning (FL) models. The set of FL solutions overviewed in this work is part of the 2021 International Telecommunication Union (ITU) AI for 5G Challenge.}
	\end{abstract}
	\ITUpar
	
	\ITUpar
	\ITUpar
	
	
	\section{Introduction} 
	\label{sec:intro}
	
	Wireless networks are evolving towards artificial intelligence (AI) / machine learning (ML)-driven systems able to address the overwhelming requirements of future mobile communications~\cite{morocho2019machine,akhtar2020shift}, namely the fifth generation (5G), beyond 5G (B5G), and the sixth generation (6G). The application of ML for networking can be found at different communication layers and parts of a network, e.g., network management to drive the self-organizing networks (SON) paradigm~\cite{klaine2017survey}, optimization of the medium access control (MAC) layer in decentralized channel access~\cite{bkassiny2012survey}, or AI-native physical communication protocols~\cite{oshea2017introduction,hoydis2021toward}. The fact is that AI/ML can leverage the vast amount of network and user data to generate new knowledge that allows improving the network performance and, hence, making progress in the development of novel network applications such as those based on extended reality.\ITUpar
	
	Nevertheless, the use of ML in communications also raises concerns of different nature. First, ML-based solutions typically require a lot of energy for training complex models (e.g., neural networks) and high bandwidth for exchanging training data, which typically needs to be centralized to a single point. Moreover, the massive usage of networking data for ML may threaten security and users' privacy. The privacy issue may be exacerbated in decentralized networks such as IEEE 802.11 wireless local area networks (WLANs), whereby the lack of a central network manager may make inter-WLAN interactions unreliable.\ITUpar 
	
	To address some of the challenges posed by traditional ML training, Federated Learning (FL) optimization was introduced in~\cite{federated1} as a distributed training paradigm that allows keeping the data at its source. Since then, a significant number of FL applications have flourished across different fields, such as medicine~\cite{federated2}, autonomous driving~\cite{federated3}, UAV-based wireless networks~\cite{federated4}. FL has become attractive to foster collaboration among different parties interested in solving a common problem. Under the management of a central server (typically, a neutral entity), FL participants contribute to building a common ML model, by sharing model weights generated using its own local data, rather than forwarding raw data for centralized training.\ITUpar
	
	In this paper, we study the application of FL models to the IEEE 802.11 spatial reuse (SR) problem, which aims to enhance spectral efficiency by adjusting the devices' carrier sense area to increase the number of concurrent transmissions in overlapping deployments. IEEE WLANs are an important part of the B5G ecosystem as it represents a cost-effective but high-performance solution for the access network. In particular, we overview the output of the problem statement entitled \textit{``ITU-ML5G-PS-004: Federated Learning for Spatial Reuse in a multi-BSS (Basic Service Set) scenario''}, which was part of the 2021 International Telecommunication Union (ITU) AI for 5G Challenge~\cite{bib2}.\footnote{The ITU AI/ML challenge is a global competition that gathers professionals, researchers, practitioners, and students from all around the globe to solve relevant problems on ML for communications.} \textcolor{black}{The purpose of the challenge was the exploration of federated solutions to predict the performance of IEEE 802.11ax (11ax) networks applying SR. Such a performance prediction solution is called to be an essential part of ML-assisted networks, which overarching goal is optimization. To address the performance prediction problem, a dataset with simulated measurements on crowded 11ax deployments applying SR was provided, which was used to develop the FL solutions presented in this paper.} The usage of simulated data for enriching training datasets is another relevant topic for enabling ML in communications~\cite{wilhelmi2021usage}.\ITUpar
	
	The main contributions of this paper are as follows:
	\begin{itemize}
		\item We overview the SR technology for both 11ax and future amendments and propose the usage of FL to address it.
		\item We provide a dataset with 11ax SR measurements for next-generation WLANs. The dataset is open and can be accessed at~\cite{dataset}. 
		\item We overview the set of FL solutions proposed by the participants of the 2021 ITU AI for 5G Challenge \textcolor{black}{to predict the performance of novel IEEE 802.11ax SR WLANs}. Table~\ref{tab:tab0} briefly summarizes the proposed models, as well as the main motivation behind them.
	\end{itemize}\ITUpar
	
	\begin{table}[!t]
		\centering
		\caption{Summary of the ML models proposed by the participants of the challenge.}\label{tab:tab0} 
		\begin{small}
			\resizebox{0.65\columnwidth}{!}{%
				\begin{tabular}{|c|c|c|c|}
					\hline
					\textbf{Team} & \textbf{Proposed Model} & \multicolumn{1}{c|}{\textbf{Motivation}} & \textbf{Ref.} \\ \hline
					FederationS & \begin{tabular}[c]{@{}l@{}}DNN with two\\parallel branches\end{tabular} & \begin{tabular}[c]{@{}l@{}}Exploit relationships among\\training features\end{tabular} & \cite{federations2021repository} \\ \hline
					\begin{tabular}[c]{@{}c@{}}FedIPC\end{tabular} & \begin{tabular}[c]{@{}c@{}}NN with a Multi-Output \\Regression Objective\end{tabular} & \begin{tabular}[c]{@{}l@{}}Take advantage of knowledge\\on wireless operation\end{tabular} &  \cite{fedipc2021repository}\\ \hline
					\begin{tabular}[c]{@{}c@{}}WirelessAI\end{tabular} & \begin{tabular}[c]{@{}c@{}} CNN with FCNN\end{tabular} & \begin{tabular}[c]{@{}l@{}} Exploit graph representations\\in wireless networks \end{tabular} & \cite{wirelessai2021repository} \\ \hline
				\end{tabular}
			}
		\end{small}
	\end{table}
	
	The rest of the paper is structured as follows. Section~\ref{section:spatial_reuse} introduces the SR problem in 11ax and future WLANs. Section~\ref{section:federated_learning} provides some basics on FL with special emphasis on its applications in networking. Section~\ref{section:dataset} overviews the provided SR dataset for training ML models. The solutions proposed by the challenge participants are described in detail in Section~\ref{section:solutions} and evaluated in Section~\ref{section:performance}. Section~\ref{section:conclusions} concludes the paper with some remarks and future directions.
	
	\section{Spatial Reuse in 802.11ax WLANs: overview and research gaps}
	\label{section:spatial_reuse}
	
	IEEE 802.11 technology, commonly known as Wi-Fi, is one of the most popular solutions for the access network due to its ease of deployment and low cost (it operates on unlicensed bands). However, its fundamental operation is based on carrier sense multiple access (CSMA), whose performance is well known to degrade when dealing with a large number of concurrent users~\cite{ziouva2002csma}. To address the issues raised by network density and to meet the increasingly strict requirements posed by next-generation applications (e.g., virtual reality), 802.11 amendments introduce novel functionalities and protocol enhancements. For instance, standards 802.11n (2009) and 802.11ac (2013) provided high throughput (HT) and very high throughput (VHT) devices by including, for instance, the application channel bonding (CB), whereby basic channels could be aggregated to increase the capacity of a single transmission.\ITUpar
	
	As for the SR operation~\cite{bib9}, it was recently introduced by the IEEE 802.11ax (2021) standard~\cite{bellalta2016ieee} to increase the number of parallel transmissions in overlapping basic service sets (OBSS). Among other features like orthogonal frequency division multiple access (OFDMA), or downlink/uplink multi-user multiple input-multiple-output (MU-MIMO), SR aims at enhancing the performance and efficiency. To do so, it provides two different operational modes:
	\begin{enumerate}
		\item OBSS Packet Detect-based SR (OBSS/PD-based SR).
		\item Parametrized Spatial Reuse (PSR).
	\end{enumerate}
	
	The main difference between the two mechanisms lies in the way SR transmission opportunities (TXOPs) are detected by devices implementing them. While OBSS/PD-based SR operates in the downlink, PSR is designed for the uplink. In what follows, we focus on OBSS/PD, which has gained more interest and is under consideration for evolution in the future amendments, such as the IEEE 802.11be (11be)~\cite{lopez2019ieee}. A comprehensive overview of these two mechanisms can be found in~\cite{bib9}.\ITUpar
	
	In essence, OBSS/PD-based SR allows devices to transmit in parallel with others that gained channel access beforehand. To do so, a new OBSS/PD threshold is defined to be applied when an incoming detected transmission marks the radio channel as busy through clear channel assessment (CCA) operation. CCA allows overlapping devices to share a common channel and is triggered when the preamble of a Wi-Fi transmission is identified. Provided that the OBSS/PD threshold allows initiating a new SR transmission, a transmit power limitation must be applied so that the generated interference does not affect the original transmission.\ITUpar
	
	\begin{figure}[!tb]
		\begin{subfigure}[b]{\linewidth}
			\centering
			\includegraphics[width=.45\textwidth]{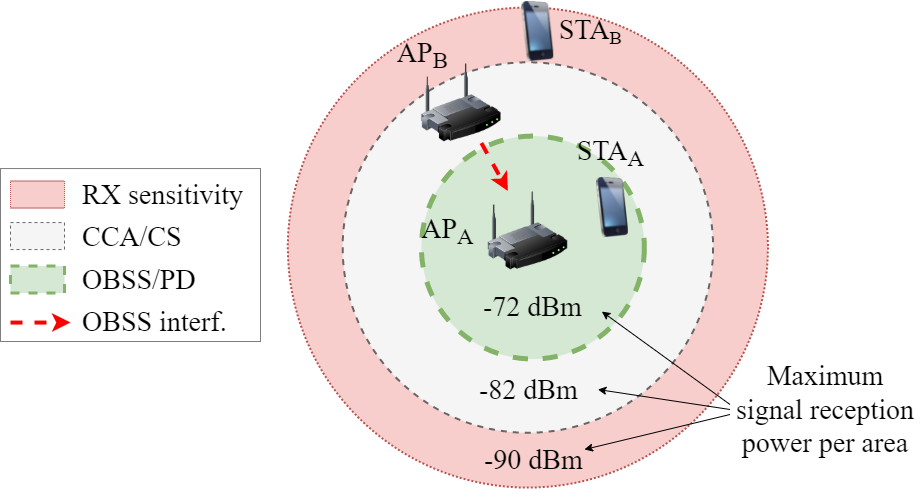}
			\caption{}
			\label{fig:obsspd_1}
		\end{subfigure}
		\begin{subfigure}[b]{\linewidth}
			\centering
			\includegraphics[width=0.45\textwidth]{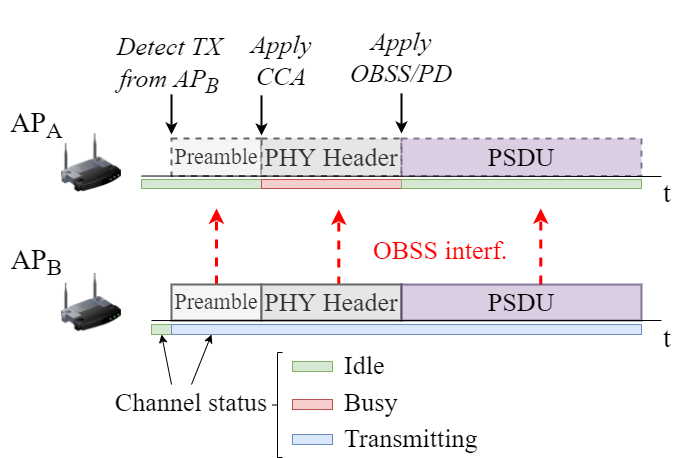}
			\caption{}
			\label{fig:obsspd_2}
		\end{subfigure}
		\caption{IEEE 802.11ax OBSS/PD-based SR operation: (a) signal reception areas, (b) diagram of packet exchange.}\label{fig:sr_operation} 
	\end{figure}
	
	The OBSS/PD SR operation is illustrated in Fig.~\ref{fig:sr_operation} for two overlapping access points (APs), AP$_\text{A}$ and AP$_\text{B}$. As shown in Fig.~\ref{fig:obsspd_1}, AP$_\text{A}$ detects the signals from all the considered devices (represented by the red area), including AP$_\text{B}$ and station B (STA$_\text{B}$), which belong to a different BSS. In particular, AP$_\text{B}$ is inside the carrier sense area of \textcolor{black}{AP$_\text{A}$} (represented by the gray area), so they both must contend for the channel whenever the other starts a transmission (e.g., to its associated STAs). Nevertheless, thanks to the OBSS/PD-based SR operation, AP$_\text{A}$ can ignore AP$_\text{B}$'s transmissions when applying the OBSS/PD threshold (represented by the green area). At the packet level (shown in Fig.~\ref{fig:obsspd_2}), AP$_\text{A}$ starts decoding the preamble of a new transmission from AP$_\text{B}$, which has initially gained the access to the medium using CSMA with collision avoidance (CSMA/CA). From the preamble reception, AP$_\text{A}$ determines that the channel is busy at the MAC layer due to the CCA operation. But, using the SR mechanism, AP$_\text{A}$ identifies an SR TXOP because the incoming signal is below the OBSS/PD threshold. Hence, AP$_\text{A}$ can initiate a transmission before AP$_\text{B}$ leaves the channel, provided that a transmit power restriction is applied, denoted by $\text{TX\_PWR}_\text{max}$, for the sake of not affecting AP$_\text{B}$'s transmission\textcolor{black}{~\cite{tgax}}:
	\begin{equation}
		\text{TX\_PWR}_\text{max} = \text{TX\_PWR}_\text{ref} - (\text{OBSS/PD} - \text{OBSS/PD}_\text{min}),	\label{eq:tx_power_limit}
	\end{equation}
	where $\text{TX\_PWR}_\text{ref}$ is the transmit power reference (set to 21 dBm or 25 dBm, depending on the device's antenna capabilities), $\text{OBSS/PD}$ is the selected OBSS/PD threshold for detecting SR TXOPs, and $\text{OBSS/PD}_\text{min}$ is the minimum OBSS/PD threshold (fixed to -82 dBm).\ITUpar
	
	While SR promises to enhance spectral efficiency in dense OBSS deployments, its actual performance is hindered by the proper selection of the OBSS/PD threshold, which may not be trivial due to the complex inter-device interactions in a WLAN. The fact is that the OBSS/PD-based SR operation is a decentralized mechanism that only considers the interactions between principal transmitters (i.e., devices gaining access to the channel for transmitting), but does not account for either the interference at the recipients of such transmissions or the impact of uplink control frames (e.g., acknowledgment packets). Since the standard does not provide any method for selecting the proper OBSS/PD threshold, there is an imperative need for finding effective mechanisms to leverage the SR operation.\ITUpar
	
	ML, in this context, is considered a promising tool to capture the complex interactions among IEEE 802.11 devices applying SR. \textcolor{black}{In general, ML has been applied to a plethora of problems in IEEE 802.11 networks, including PHY optimization (rate selection~\cite{karmakar2016dynamic}, resource allocation~\cite{testolin2014machine}), assisting management operations (e.g., AP selection and handover~\cite{wu2020novel}, channel band selection~\cite{niyato2009approach}), or supporting novel features like MU-MIMO or channel bonding with enhanced monitoring, analytics, and decision-making~\cite{karmakar2019intelligent, barrachina2021multi}. For further details on ML application to Wi-Fi, we refer the interested reader to the comprehensive survey in~\cite{szott2021wifi}.}\ITUpar
	
	\textcolor{black}{In the particular case of SR,} most of the literature has so far focused on reinforcement learning (RL) and online learning techniques, whereby agents attempt to learn the best OBSS/PD configuration sequentially. In~\cite{wilhelmi2019collaborative, wilhelmi2019potential}, the authors modeled the decentralized SR problem as multi-armed bandits (MAB), an online learning framework whereby agents attempt to address the exploration-exploitation trade-off. While~\cite{wilhelmi2019collaborative} studied the problem by using selfish rewards in a competitive environment,~\cite{wilhelmi2019potential} considered shared rewards for the sake of maximizing fairness. Other RL-based approaches can be found in~\cite{bardou2021improving} and~\cite{yin2019learning}.\ITUpar
	
	The online learning paradigm turns out to be a cost-effective solution to the decentralized SR problem thanks to its ability for solving complex partial information problems. In addition, WLANs typically experience a high variability both in terms of devices' mobility and activation/deactivation, so past learned information may become easily outdated. However, as shown in~\cite{wilhelmi2019potential}, online learning may have some pitfalls when applied to dense WLANs, mainly raised by the high action-decision space, the non-stationarity of agents' rewards in competitive settings, or the complexity of finding a proper shared reward that enables maximizing the overall network performance.\ITUpar
	
	For those reasons, in this paper, we focus on the suitability of supervised learning methods, mostly based on deep learning (DL), for the SR problem in WLANs. To the best of our knowledge, this approach has not  been studied before in the context of SR. A centralized DL-based method was proposed in~\cite{jamil2016novel} to jointly select the transmission power and the CCA, but not in the context of 11ax SR operation. DL was also applied in~\cite{soto2021atari} to address the channel bonding problem in dense WLANs. This and other DL solutions for the dynamic channel bonding problem in IEEE 802.11ax WLANs were overviewed in~\cite{wilhelmi2021machine}.\ITUpar
	
	We note that the overviewed works on DL consider centralized approaches, which require data to be gathered at a single point for training a static model, which is then used homogeneously across all the AI-enabled devices. Nevertheless, in practice, some deployments (e.g., residential WLANs) may have limitations in terms of computation, storage, or communication capabilities (for instance, low-throughput connections, intermittent availability). Moreover, separate WLAN deployments can be substantially different, thus requiring specialized models (rather than general ones). To address these limitations of centralized learning on heterogeneous deployments, we focus on the FL paradigm, introduced in the next section.\ITUpar
	
	\section{An Introduction to Federated Learning for Networking}
	\label{section:federated_learning}
	
	The FL optimization paradigm was first introduced in~\cite{konevcny2016federated} to address some critical issues of traditional centralized ML mechanisms. In FL, training is done at end devices (or clients), which do not share their training data with others. Instead, ML model updates are provided and aggregated under the management of a typically central server. By removing the procedures related to data exchange, FL decreases the communication overhead and enhances user privacy and security. Moreover, FL is an appealing solution for dealing with heterogeneous sets of clients, thus allowing to create specialized models according to clients' characteristics. With that, FL has the potential to revolutionize ML implementations, bringing them closer to practical applications and use cases. Many examples of FL have emerged in recent years, including, but not limited to, in medicine~\cite{nguyen2021federated2}, finance~\cite{long2020federated}, industry 4.0~\cite{que2020blockchain}, or telecommunications~\cite{amiri2020federated,lim2020federated}.\ITUpar
	
	In the telecommunications realm, novel ML solutions require handling a vast amount of data, often highly distributed across the network. These kinds of resource-demanding applications may threaten the stability of the network on the one hand and may experience low performance due to the communication bottleneck on the other hand. FL can potentially alleviate some of these issues by reducing the overheads generated by the ML operation while providing good performance. FL also contributes to enhancing privacy, which is a critical issue in communications. FL applications in communications~\cite{yang2021federated} include autonomous driving~\cite{li2021privacy}, unmanned aerial vehicle (UAV)-based wireless networks~\cite{brik2020federated}, edge computing~\cite{wang2019inedge}, physical layer optimization~\cite{mashhadi2021federated}, or Internet-of-Things (IoT) intelligence~\cite{khan2021federated}.\ITUpar
	
\begin{figure}[!tb]
		\centering
		\includegraphics[width=.6\columnwidth]{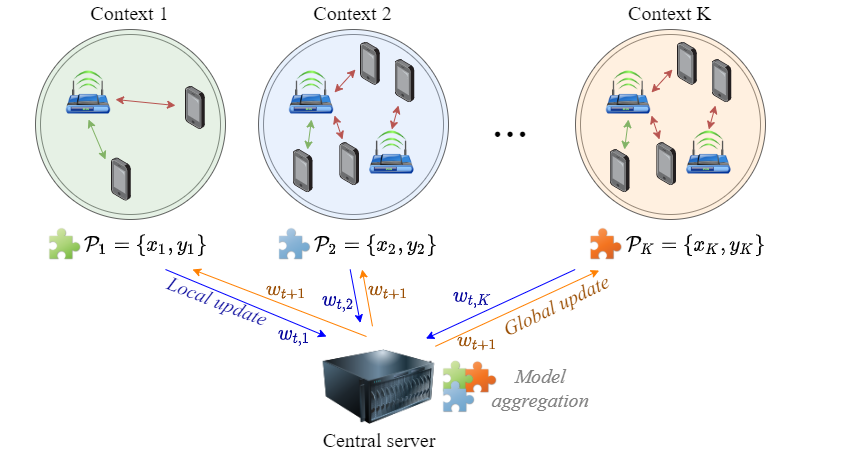}
		\caption{\textcolor{black}{FL operation with WLAN contexts.}}\label{fig:federated_optimization}
	\end{figure}

	The generic FL algorithm operates iteratively, generating a global model update at each iteration with the help of a subset of clients. Each algorithm's iteration follows the following general steps (see Fig.~\ref{fig:federated_optimization}): 
	\begin{enumerate}
		\item A set of $\mathcal{K}=\{1,2,...,K\}$ clients download the current model parameters, $w_t$, from the central server (also called the parameter server).
		\item Clients perform training in parallel using their local datasets $\mathcal{D}^{(k)}$ (with size $N^{(k)}$) and update the model weights accordingly, denoted by $w^{(k)}\in \mathbb{R}^d$.
		\item The server pulls the model updates from the participating clients (a subset of clients may be selected in each FL round for the sake of performance) and orchestrates weight aggregation to generate an updated global model $w_{t+1}\in \mathbb{R}^d$.
		\item Above steps are repeated until convergence, i.e., until a time horizon is completed or a certain accuracy goal is met. 
	\end{enumerate} 
	
	At this point, it is important to highlight the federated averaging (FedAvg) method~\cite{mcmahan2017communication}, which is based on stochastic gradient descent (SGD) optimization and performs well for non-convex problems. In FedAvg (shown in Alg.~\ref{alg:fl}), clients perform several batch updates at each iteration using local data to update the global model parameters. Unlike in classical federated stochastic gradient descent (FedSGD), where gradients are exchanged, FedAvg considers sharing model updates (e.g., the parameters of a neural network). By applying multiple rounds of training, FL seeks to minimize a global finite-sum cost function $l(w)$ by optimizing the global model parameters $w$:
	\begin{equation}
		\min_{w \in \mathbb{R}^d} l(w) = \min_{w\in \mathbb{R}^d} \sum_{k=1}^{K} \frac{N^{(k)}}{N} l^{(k)}(w,\mathcal{D}^{(k)}),
		\label{eq:fl_loss_goal}
	\end{equation}
	where $l^{(k)}(w,\mathcal{D}^{(k)})$ is the loss experienced by client $k$ when using the global model $w$ on its local data, and $N$ is the total size of the distributed dataset, i.e., $N = \sum_{\forall k \in \mathcal{K}} N^{(k)}$.\ITUpar
	
	To compute local updates, clients run $E$ epochs of SGD based on the target local loss function $l^{(k)}$ and the batch size $B$ applied to local data $\mathcal{D}^{(k)}$. Using a learning rate $\eta$, local updates are obtained by:
	\begin{equation}
		w_{t+1}^{(k)} \leftarrow w_{t}^{(k)} - \eta \nabla l^{(k)}(w_t,\mathcal{D}^{(k)})
		\label{eq:fl_local_update}
	\end{equation}
	
	Finally, being $\eta$ the learning rate, the server aggregates clients' weights based on the importance $\alpha_k$ assigned to each client which may be set according to local dataset lengths (as indicated in Eq.~\eqref{eq:fl_loss_goal}):
	\begin{equation}
		w_{t+1} = \sum_{k=1}^{K} \alpha^{(k)} w_{t+1}^{(k)}  
		\label{eq:fl_aggregation}
	\end{equation}
	
	\begin{algorithm}[t!]
		\caption{Federated Averaging (FedAvg) }\label{alg:fl}
		\begin{algorithmic}[1]
			\For{$t=1,2,\ldots,T$}
			\For{$k \in \mathbb{K}_{tr}$} in parallel
			\State Pull $\boldsymbol{w}_{t}$ from central server: $\boldsymbol{w}^{(k)}_{t,0}=\boldsymbol{w}_{t}$
			\For{$e=1,\ldots,E$}
			\State Update model: $\boldsymbol{w}^{(k)}_{t,e}=\boldsymbol{w}_{t,e}^{(k)}-\eta_{t}\nabla l^{(k)}_{t,e}$
			\EndFor
			\State Push $\boldsymbol{w}_{t+1}^{(k)} \leftarrow \boldsymbol{w}_{t,E}^{(k)}$
			\EndFor
			\State{\textbf{FedAvg}:} $\boldsymbol{w}_{t+1}=\frac{1}{ \vert \mathbb{K}_{tr} \vert}\sum_{k\in\mathbb{K}_{tr}} \boldsymbol{w}_{t+1}^{(k)}$
			\EndFor
		\end{algorithmic}
	\end{algorithm}
	
	Beyond FedAvg, other optimization mechanisms have been proposed to improve the convergence and efficiency of FL~\cite{reddi2020adaptive, ozfatura2020accelerated}. For further details on FL, we refer the interested reader to the works in~\cite{zhang2021survey, li2020federated}, and to~\cite{chen2021distributed} for the implementations of FL over wireless networks in particular.\ITUpar
	
	\section{Open simulated dataset on IEEE 802.11ax SR}
	\label{section:dataset}
	
	Supervised ML methods typically require a significant amount of high-quality data to perform well. Training data is usually obtained either from network activity~\cite{turkcelldataset} or from measurement campaigns~\cite{barrachina2021wifi}. However, obtaining real traces from networks can be challenging due to proprietary limitations (data owners are reluctant to share their assets), data privacy issues (most network data is generated by final users), or difficulties in obtaining data from a rich set of situations (anomalies are hard to reproduce and identify). In this sense, the usage of synthetic datasets for model training is gaining attention~\cite{wilhelmi2021usage}. Such datasets can be obtained, for instance, from network simulators (e.g., ns-3, OMNET++, OPNET). Simulators are a cost-effective solution for generating comprehensive datasets. Some prominent examples of synthetic datasets oriented to ML training can be found in~\cite{suarez2021graph, dataset2020}.\ITUpar
	
	As for the provided dataset on 11ax SR~\cite{dataset}, it has been generated with Komondor~\cite{bib8}, an open-source IEEE 802.11ax-oriented simulator that includes features like channel bonding or SR. Komondor does not implement the targeted functionalities, but its execution is also lightweight, thus allowing for generating large datasets corresponding to massive WLAN deployments.\ITUpar
	
\begin{figure}[!tb]
		\centering
		\includegraphics[width=.4\columnwidth]{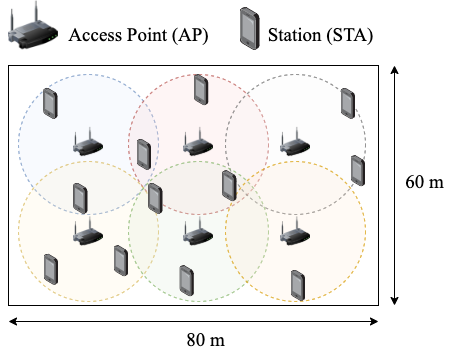}
		\caption{Example of a simulated WLAN deployment.}\label{fig:simulation_scenario} 
	\end{figure}

	The dataset contains both training and test files, which include the results obtained from several simulated random deployments applying 11ax SR (see the example random deployment in Fig.~\ref{fig:simulation_scenario}). More specifically, a set of three baseline scenarios was considered to represent different types of deployments. Considering the features in each type of scenario (e.g., maximum number of STAs per BSS, minimum distance between APs), 1,000 random deployments of each type were generated for training. Each simulated deployment corresponds to a context $k\in \mathbb{K}$, where the BSS of interest (namely, BSS$_\text{A}$) is used as a client for FL optimization. To enrich contexts with data, each BSS includes information for each possible OBSS/PD configuration $\tau$ (i.e., from -82 dBm to -62 dBm with 1 dBm precision). Finally, for the test dataset, 1,000 more deployments were simulated using more relaxed constraints. In this case, a single random OBSS/PD configuration was selected in each deployment. Table~\ref{tab:tab1} provides an overview of the entire dataset.\ITUpar
	
	\begin{table}
		\centering
		\caption{Summary of the scenarios of the dataset.}\label{tab:tab1} 
		\resizebox{.6\linewidth}{!}{\begin{tabular}{|c|c|c|c|c|c|}
				\hline
				& \textbf{Sce id} & \textbf{Num. APs} & \textbf{Num. STAs} & \textbf{d\_min(APs)} & \textbf{\begin{tabular}[c]{@{}c@{}}Context \\ variations\end{tabular}} \\ \hline
				\multirow{3}{*}{\textbf{Training}} & training1 & \multirow{4}{*}{2-6} & 1 & 10 m & None \\ \cline{2-2} \cline{4-6} 
				& training2 &  & 1-4 & 10 m & None \\ \cline{2-2} \cline{4-6} 
				& training3 &  & 1-4 & None & \begin{tabular}[c]{@{}c@{}}Up to 20 \\ locations\end{tabular} \\ \cline{1-2} \cline{4-6} 
				\textbf{Test} & test &  & 2-4 & None & None \\ \hline
		\end{tabular}}
	\end{table}
	
	Training scenario \textit{training1} considers BSSs with only one STA, which is useful to minimize the impact of uplink transmissions, thus allowing to focus on inter-AP interactions only. In contrast, scenarios \textit{training2} and \textit{training3} consider up to 4 STAs per AP, which contribute generating more traffic in the uplink. As for the minimum distance between APs (d$_\text{min}$), it is set to 10~m in scenarios \textit{training1} and \textit{training2}, whereas the rest have no limitation. Furthermore, contexts in scenario \textit{training3} contain richer datasets by simulating variations of the same deployments using different STA locations. \ITUpar
	
	The information included in simulated files is divided into features and label. Concerning features for training, we find the following key elements:
	\begin{enumerate}
		\item \textbf{Type of node:} indicates whether the node is an AP or an STA.
		\item \textbf{BSS id:} identifier of the BSS to which the node belongs.
		\item \textbf{Node location:} \{x,y,z\} position of nodes in the map.
		\item \textbf{Primary channel:} main frequency channel used for transmitting and for carrier sensing.
		\item \textbf{Transmit power:} default transmit power used for transmitting frames.
		\item \textbf{OBSS/PD threshold:} sensitivity used within the OBSS/PD-based SR operation.
		\item \textbf{Received signal strength indicator (RSSI):} average signal quality experienced by STAs during reception phases.
		\item \textbf{Inter-BSS interference:} average power sensed from devices belonging to other BSSs.
		\item \textbf{Signal-to-interference-plus-noise ratio (SINR):} average SINR experienced by STAs when receiving data from their AP.
	\end{enumerate}
	
	\begin{figure*}[t!]
		\centering
		\includegraphics[width=.78\linewidth]{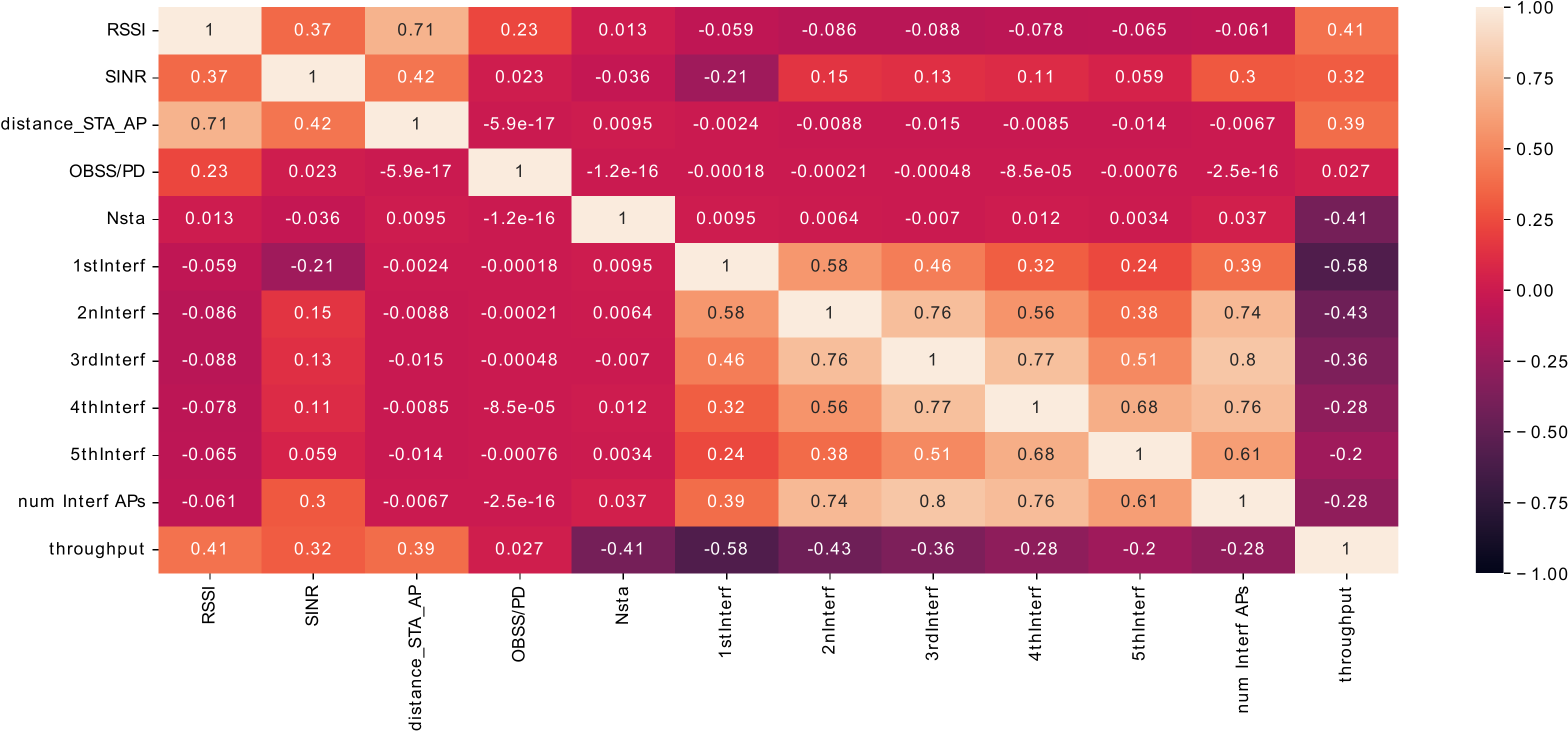}
		\caption{Correlation between different input and output variables of the dataset.}
		\label{fig:Correlation heatmap}
	\end{figure*}
	
	It is worth noting that most of the extracted information is typically obtained on a continuous basis in a real system. Indeed, the RSSI, SINR and throughput measurements can be reported periodically by STAs. Interference powers can be measured during the listen-before-transmit (LBT) phase at the AP, employing multi-antenna processing techniques to separate the different interfering sources. In addition, time-of-arrival (TOA) ranging techniques can determine the distance between STAs and APs.\ITUpar
	
	As for the label, we provide the throughput $\gamma^{(k)}_{j,\tau}$ obtained by each STA $j$ in context $k$ during the simulation, provided that the OBSS/PD configuration $\tau$ is used. Predicting the throughput is the goal of the implemented FL solutions described in the next section. Notice, as well, that other Key Performance Indicators (KPIs) such as the average delay or the number of SR TXOPs could have been considered.\ITUpar
	
\begin{figure}[tb!]
		\centering
		\includegraphics[width=.425\columnwidth]{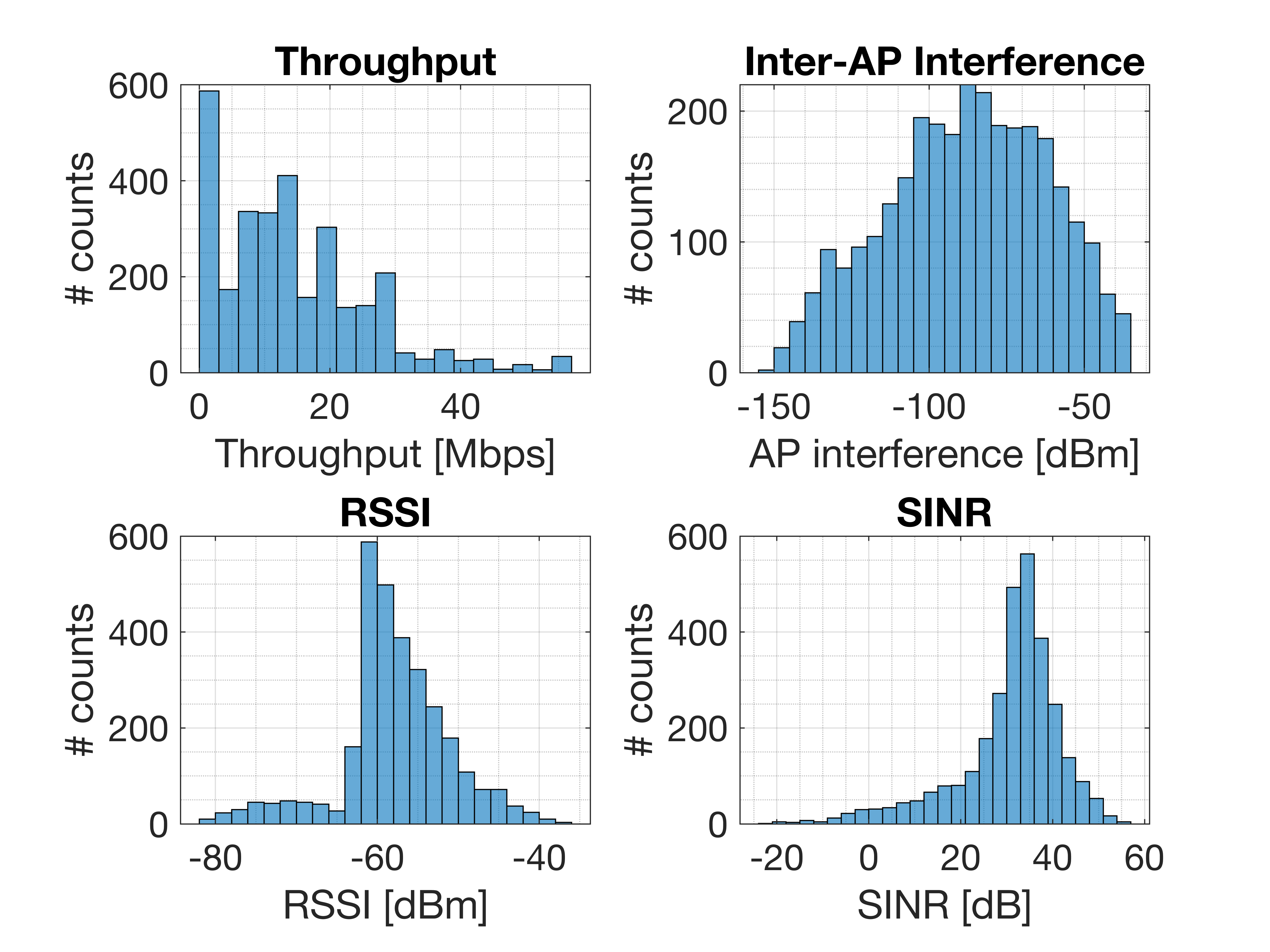}
		\caption{Histogram of relevant features from the dataset.}
		\label{fig:hist_features}
	\end{figure}

	To conclude this section, we show the correlation matrix between input and output variables in Fig.~\ref{fig:Correlation heatmap}, which is later used as a motivation for some of the proposed ML solutions. Correlation values close to $0$ indicate a lack of relations and structure between the data corresponding to these variables, while correlation values close to $-1$ and $+1$ indicate a perfect negative and positive correlation between variables, respectively. Finally, Fig.~\ref{fig:hist_features} shows the histogram of some of the most relevant features from the entire dataset.
	
	\section{Federated Learning Solutions for Spatial Reuse}
	\label{section:solutions}
	
	In this section, we describe the solutions proposed by the participants of the 2021 ITU AI for 5G Challenge: \textit{FederationS}, \textit{FedIPC}, and \textit{WirelessAI}.
	
	\subsection{FederationS}
	
	
	\textcolor{black}{This} solution is designed in three stages. In the first stage, we analyze and pre-process available datasets. In the second stage, using gained insights from the data analysis, we define a deep neural network (DNN) model running in each client. In the final stage, we describe the proposed FL algorithm. \ITUpar
	
	In the data analysis stage, we consider scenarios \textit{training2} and \textit{training3}, containing $2000$ different IEEE 802.11ax deployments (see Table~\ref{tab:tab1} for further details). We extract several features available in the simulator’s output files from these scenarios, namely the OBSS/PD configuration, the RSSI, the interference at the reference AP from other APs, the SINR, and the throughput of each STA. We also obtain additional information using available data in the simulator’s input files. In particular, we extract the coordinates of APs and STAs to compute the Euclidean distances among them. In addition, we obtain the number of STA served by the reference AP and the number of interfering APs.\ITUpar
	
	To obtain the final dataset to be used to train our model, we pre-process the data through different steps. First, we clean the input and output data parsed from the simulator files removing all non-numerical values from the dataset. Then, we arrange the data of each STA to form 1-D vectors with $11$ numerical entries used as the input of the model and containing all measurements and system parameters. Conversely, we define the STA throughput as the target variable and output of the model. To note that the features present entirely different ranges between maximum and minimum values and are expressed with different units of measurements, e.g. dBm for RSSI and interference power, dB for SINR, and meters for distances. To balance each feature's contribution to the overall model predictions, we re-scale the features with the Min-Max normalization method that transforms all features' in the range $[0,1]$. Finally, when input data are missing, like when the number of interfering AP reported is less than the minimum recorded according to our system settings, we assign those values with $0$s. The activation function that we will explain later is chosen to keep neurons inactive when $0$s are present at the input.\ITUpar
	
	
	
	To decide the ML method to be used, we make the following two main observations from the correlation analysis done in Fig.~\ref{fig:Correlation heatmap}:
	\begin{enumerate}
		\item Most features show a strong positive or negative correlation with the output variable (throughput). As expected, only the OBSS/PD feature does not directly affect the throughput. Otherwise, the problem would be trivial to model. Indeed, the OBSS/PD value is correlated to the RSSI, which affects SINR and throughput variables, showing that relationships between input and output values of the system are not straightforward to characterize with domain-based models. This justifies the adoption of DNN, which are extensively used for their capabilities to model nonlinear relationships. 
		\item Two regions depicted with lighter colors at the top left and at the bottom right of the correlation matrix identify two groups of features that show a strong positive correlation between input variables. Thus, in the DNN architecture design, these inputs of the model need to be fully connected. In contrast, the two regions at the top right and bottom left are characterized by elements with close to zero correlation values, meaning that the relationship between features is weak. Therefore, these connections are expected to bring a low contribution to the predictions and can be dropped in the DNN architecture design. 
	\end{enumerate}
	
\begin{figure}[tb!]
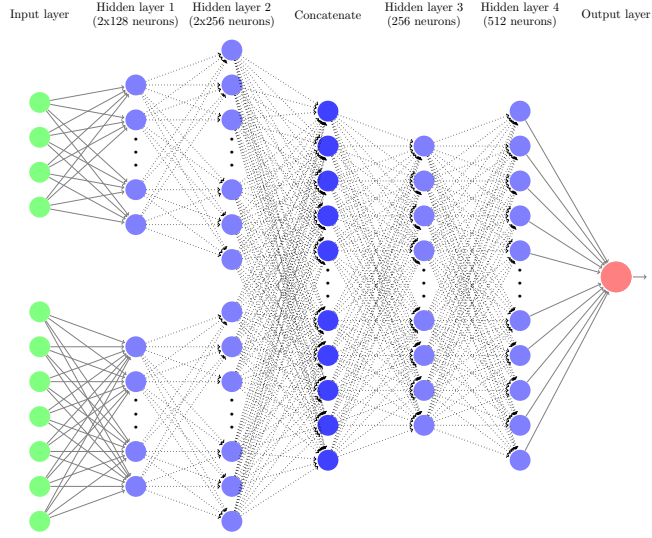

		\centering
		\includestandalone[width=.5\linewidth]{tikz_figures/ann_small}
		\caption{Visualization of FederationS' DNN structure.}
		\label{fig:DNN model}
	\end{figure}

	Based on this, in the following design stage, we model DNN architecture as represented in Fig.~\ref{fig:DNN model}. First, we split the DNN model into two parallel branches. The inputs of the first branch are the features constituting the first block, i.e. RSSI, SINR, distance STA-AP, and OBSS-PD threshold. At the same time, features like the number of STA, the power received from interfering AP, and the number of interfering AP form the second block of features and are used as input of the second branch. The input layers are followed by two hidden layers defined for each branch separately. We use a concatenation layer to merge the output of these two branches. The result of the concatenation is then used as input of two additional hidden layers, which are connected to the output layer of the model. We adopt the hyperbolic tangent (\emph{tanh}) activation function to provide positive and negative outputs and keep neurons inactive when the inputs are $0$s. Finally, we add dropout layers after each layer before the output layer to reduce overfitting.\ITUpar
	
	As for the FL solution, it is based on the implementation outlined in~\cite{federated_repo}. \textcolor{black}{However, the FederationS algorithm combines the trained weights in a novel way, which is designed specifically for the problem of performance prediction in WLANs. Moreover, the aggregation process at the central server is tailored to maximise the gain from contexts with more data samples.} In particular, our proposed FL solution follows the steps described in Algorithm~\ref{alg:fed}.\ITUpar 
	
	
	\begin{algorithm}[tb!]
		\caption{FederationS solution.}
		\begin{algorithmic}[1]
			\State Init set $\mathbb{K}$, i.e., init $K$ contexts with data samples
			\State From $\mathbb{K}$, select $N_{eval}$ to create $\mathbb{K}_{val}$
			\State Create a new set of contexts $\mathbb{K}_{tr} =  \mathbb{K} \cap \mathbb{K}_{val}$
			\State Server initializes model parameters $\theta_0$ and $W_0$
			\State The server transmits $\theta^{(k)}_0, w^{(k)}_0$ to $k$-th contexts
			\For{communication epoch $t=1,2,...,T$}
			\State Rand. select $N_{tr}$ contexts from $\mathbb{K}_{tr}$ to get $\mathbb{K}_{ep}$
			\For{$i$-th context in $\mathbb{K}_{ep}$}
			\State Split samples in $\beta$ ($\frac{n_i}{B}$ batches of size $B$)
			\State Where $n^{(i)}$ is the number of data samples 
			\For{batch $b$ in $\beta$}
			\State  $\theta^{(i)}_t \leftarrow \theta^{(i)}_{t-1}-\eta \nabla l(\theta^{(i)}_{t-1};b)$
			\State  $w^{(i)}_t \leftarrow  w^{(i)}_{t-1}-\eta \nabla l( w^{(i)}_{t-1};b)$
			\EndFor
			\State Determine weight $\alpha^{(i)} = n^{(i)}/N_{STA}$
			\State Transmit $\theta^{(i)}_t$, $w^{(i)}_t$, and $\alpha^{(i)}$ to central server
			\EndFor
			\State Calculate data samples weight $\alpha_{t} = \sum^{N_{tr}}_{i=1} \alpha^{(i)}$
			\State Update model: \newline 
			$\phantom{AAAA}$ $\theta^{(k)}_t = \sum^{N_{tr}}_{i=1}  \frac{\alpha^{(i)} *\theta^{(i)}_t }{\alpha_t}$, $w^{(k)}_t = \sum^{N_{tr}}_{i=1}  \frac{\alpha^{(i)} w^{(i)}_t }{\alpha_t}$
			\State The server transmits $\theta_k^t, w^{(k)}_t$ to $k$-th contexts
			\EndFor
			\State \textbf{Output:} $\theta_T$ and $w_T$
		\end{algorithmic} \label{alg:fed}
	\end{algorithm}
	
	In steps 1-5, we initialize the contexts and split them into train and validation sets. After the initialization, the training takes place locally in a subset of $N_{tr}$ contexts, which are selected randomly at every communication epoch. The training results, i.e., the trained neural network $\theta^{(i)}$ and its weights $w^{(i)}$ along with the number of data sample $n^{(i)}$, are then transmitted from the $N_{tr}$ contexts to the central server for aggregation. After the aggregation, the central server sends back the global trained model to every context, which updates their local DNN models. The training cycle repeats for $T$ communication epochs. \ITUpar
	
	One of the most important aspects of the FL training consists of aggregating the weights at the central server. Initially, we weighted the updates from each context equally, but such an approach resulted in a skewed performance toward contexts with more STA. Such behavior can be attributed to the fact that contexts with four STA have twice as many samples as contexts with only two STA. Instead, our solution proposed to weight each context update based on the number of data samples used for the training in each context \textcolor{black}{is normalized} by the number of STA in the contexts. We denote this normalization weights with $\alpha$. This approach improves the accuracy of the predictions of the throughput. \ITUpar
	
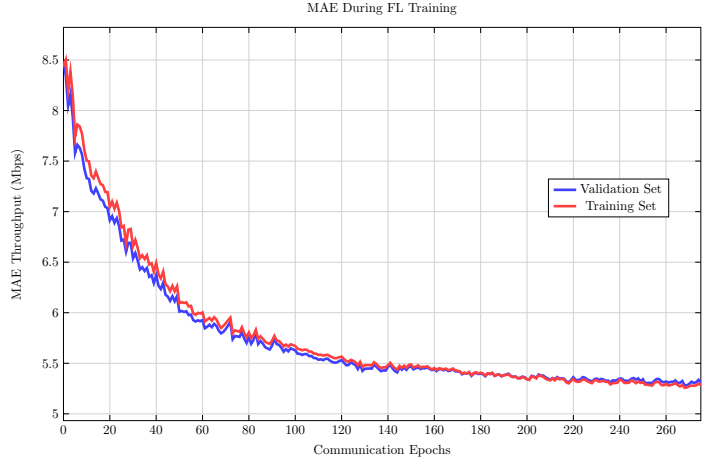
\begin{figure}[tb!]
		\centering
		\begin{tikzpicture}
\begin{axis}[
grid=both,
minor grid style={gray!35},
major grid style={gray!35},
title = MAE During FL Training,
xlabel=Communication Epochs,
height=0.65\linewidth,
width=1\linewidth,
ylabel= MAE Throughput (Mbps),
legend style={at={(0.95,0.615)}},
xmin=0,xmax=275,
every tick/.style={
	black,
	semithick,
}]

\addplot[blue!75, line width=2pt] 
table[x=round,y=mae_val,col sep=comma]{csv_data/rmse_mae_for_report.csv};
\addplot[red!75, line width=2pt] 
table[x=round,y=mae_train,col sep=comma]{csv_data/rmse_mae_for_report.csv};

\addlegendentry{Validation Set};
\addlegendentry{Training Set};
\end{axis}

\end{tikzpicture}
		\caption{MAE obtained by FederationS' over the number of communication epochs.}
		\label{fig:MAE}
	\end{figure}

	To evaluate the performance of the proposed solution (shown in Fig.~\ref{fig:MAE}), we consider the mean average error (MAE). In particular, we use a neural network trained for $T=250$ communication rounds. In Table \ref{table:parameters}, we report the list of hyper-parameters used in the submitted solution and related pre-trained DNN can be found in the GitHub repository \cite{federations2021repository}. We perform the validation using five percent of available contexts, randomly sampled from $\mathbb{K}$. In total, we had $K=1946$ available contexts. Five percent of available contexts are used for evaluation, i.e., $N_{eval}=97$. At each communication round, we select 500 contexts randomly to perform training on, i.e., $N_{tr}=500$. Furthermore, the contexts we use during the training and validation set are kept separated to prevent the data leakage and to recognize when the solution starts to over-fitting to the training data, i.e., $\mathbb{K}_{val}$,  where $\mathbb{K}_{val}  \cap \mathbb{K}_{tr} = \emptyset$.\ITUpar
	
	As shown in Fig.~\ref{fig:MAE}, the MAE decreases over the number of communication epochs. The same trend occurs for the contexts that we use during the training process (Training set) as well as for the contexts that we use only for validation (Validation set). However, the validation throughput decrease is noisier as it sometimes increases between two consecutive communication epochs. Such behavior is due to the random sampling approach as not all randomly selected sets $\mathbb{K}_{ep}$ wholesomely represent the system.\ITUpar
	
	
	\begin{table}
		\centering
		\caption{FederationS hyper-parameters}
		\begin{tabular}{|c|c|c|}
			\hline
			\multirow{5}{*}{\begin{tabular}[c]{@{}c@{}} Neural Network\\ training \\ options \end{tabular}} & Solver & Adam \cite{2014arXiv1412.6980K} \\ \cline{2-3} 
			& Batch size ($B$) & $21$ \\ \cline{2-3} 
			& Dropout & $10\%$ \\ \cline{2-3} 
			& Learning rate & $10^{-4}$ \\ \cline{2-3} 
			& L2 regularization & $10^{-5}$ \\ \hline
		\end{tabular}
		\label{table:parameters}
	\end{table}
	
	\subsection{FedIPC}
	
	We design an NN model that predicts the throughput of the STAs of a given BSS for a chosen SR configuration. The goal of this solution is to find the optimal OBSS/PD threshold maximizing the network throughput. To this end, we aim to design a neural network model which predicts the throughput of each STA in the given BSS for a chosen OBSS/PD threshold from a certain range, thus one can tune the OBSS/PD threshold by using NN architecture. In the federated learning setting, we model each context as a node, where their data consist of simulations with different thresholds. We assume these nodes cannot communicate with each other, but communicate with a parameter server in rounds, which aggregates the weights of the nodes to update the global model. Then, the parameter server distributes the updated global model to the clients.\ITUpar
	
	
	
	To train the NN model we employ the federated learning framework in the following way; we call a Wi-Fi deployment with specific characteristics such as node locations and number of interfering BSSs as a context. We consider $n$ contexts in total, where for each context, we have $s_k$ STAs per AP for context $k$. We also define $a$ as the maximum number of access point and $b$ as the maximum number of STAs per AP. Note that all contexts may have different number of STAs per AP. We also have interference sensed by APs, RSSI of the STAs assigned to $\mathrm{AP}_\mathrm{A}$, and the average SINR of each STA in $\mathrm{BSS}_A$. We can \textcolor{black}{control} threshold $\tau_k \in \{-82, -81, \ldots, -62\}$. \textcolor{black}{To simplify the problem,} we can only change the threshold of the $\mathrm{BSS}_A$, and all other BSSs' thresholds are fixed to -82 dBm. Thus, we only consider the STAs in $\mathrm{BSS}_A$. Let $\gamma_{j, \tau_{k}}^{(k)} \in \mathbb{R}$ be the throughput of $j^\mathrm{th}$ STA in the $\mathrm{BSS}_A$ of the $k^\mathrm{th}$ context, where threshold of the $\mathrm{BSS}_A$ is chosen as $\tau_k$. For context i, our objective is to find $\tau_k$ that maximizes the throughput for all STAs in the $\mathrm{BSS}_A$ of context $k$, i.e,
	\begin{equation}
		\argmax_{\vec{\tau}} \sum_{k=1}^{n} \sum_{j=1}^{s_k} \gamma_{j, \tau_{k}}^{(k)},
		\label{eq:thresholdeq}
	\end{equation}
	where $\vec{\tau} = \left[\tau_1, \tau_2, \ldots, \tau_n \right]^T$, $s_k$ is the number of STAs connected to each AP (or the number of STAs in each BSS) in context $k$. Having the knowledge of throughput values $\gamma_{j, \tau^{'}}^{(k)}$, $\forall k,j,\tau_{'}$, which is not likely, one can easily calculate $\vec{\tau}$ using~\eqref{eq:thresholdeq}. Thus, to determine the best threshold for each context $k$, we  estimate $\gamma_{j, \tau^{'}}^{(k)}$ via $\hat{\gamma}_{j, \tau_{i}}^{(k)}$ for all STA $j$ and threshold $\tau^{'}$ combinations.\ITUpar
	
	Since we cannot directly calculate or know the throughput $\gamma_{j, \tau}^{(k)}$, we estimate it via a model $\hat{\gamma}^{(k)}_{j, \tau} = f^{(k)}_{j,\tau}$, where $i$ is the context index, $j$ is the index of STA connected to $\mathrm{AP}_\mathrm{A}$ and $\tau$ is the threshold. Moreover, estimating one STA's throughput is highly related to estimating another STA's throughput in the same context.  Thus, to exploit this relation, we formulate the throughput regression problem as multi-output regression, as the following:
	\begin{equation}
		\vec{f}^{(k)}_\tau (\vec{x}_\tau^{(k)}, \vec{W}_i ) = \left[ f^{(k)}_{1,\tau} \, \, f^{(k)}_{2,\tau} \ldots f^{(k)}_{b,\tau} \right]^T,
	\end{equation}
	where $k$ is the context index, $\vec{x}_\tau^{(k)}$ is the input vector and $\vec{W}_i$ is the neural network weights of the model at context $k$. The input vector $\vec{x}_\tau^{(k)} \in \mathbb{R}^{4b+a}$ includes each STA's features in order (for STAs in $\mathrm{BSS}_A$). Each STA's features are as the following: interference sensed by APs, RSSI, the average SINR and the threshold, respectively. \textcolor{black}{When a context has less than $b$ STA per AP, we zero pad for the remaining places until the vector reaches the maximum in the dataset}. This is possible since $a$  (the maximum number of APs) and $b$ (the maximum number of STAs per AP) are fixed.\ITUpar
	
	Since every context may have different number of STAs per AP, we mask the nonexistent STAs as the following:
	\begin{equation*}
		f^{(k)}_{k, \tau} = \hat{\gamma}^{(k)}_{k, \tau} = \gamma^{(k)}_{k, \tau} = 0 , \, \, \, \, \forall k \in \{s_k + 1, \, \ldots , b \} .
	\end{equation*}
	This way, we do not backpropagate any loss for nonexistent STAs, and \textcolor{black}{the model becomes suitable for variable number of STAs per AP for every context}. Then, we define the ground truth vector as:
	\begin{equation*}
		\vec{\gamma}^{(k)}_\tau = \left[ \gamma_{1, \tau}^{(k)} \, \, \gamma_{2, \tau}^{(k)} \, \,  \ldots \, \, \gamma_{b, \tau}^{(k)} \right]^T
	\end{equation*}
	
	For the context $k$ (local node), our objective is to minimize mean-squared error for regression task for any $(\vec{x}_\tau^{(k)}, \vec{\gamma}^{(k)}_\tau)$ data point among all contexts, i.e.,
	\begin{equation*}
		\argmin_{\vec{W}_k} \sum_{\forall \tau,k}  \norm{\vec{f}^{(k)}_\tau (\vec{x}_\tau^{(k)}, \vec{W}_i ) - \vec{\gamma}^{(k)}_\tau}_2^2.
	\end{equation*}
	
	We use a feed-forward neural network with one hidden layer as our model $\vec{f}^{(k)}_\tau (\vec{x}_\tau^{(k)}, \vec{W}_k)$ with weights $\vec{W}_k = \left[  \vec{W}_k^{(1)} \, \, \vec{W}_k^{(2)} \right]$, where $\vec{f}^{(k)}_\tau (\vec{x}_\tau^{(k)}, \vec{W}_i) = \vec{W}_k^{(2)} \mathrm{ReLU} ( \vec{W}_k^{(1)}\vec{x}_\tau^{(k)}) $, $\vec{W}_k^{(2)} \in \mathbb{R}^{b \times h}$ and  $\vec{W}_k^{(1)} \in \mathbb{R}^{h \times (a+3b)}$. As seen, we use rectified linear unit (ReLU) as our activation function. Note that this neural network can easily be generalized to a neural network with multiple hidden layers, but in our case, the neural network with only 1 hidden layer \textcolor{black}{has} worked the best on the validation set.\ITUpar
	
	
	To train the proposed NN architecture under the FL paradigm, FedAvg is applied (see Section~\ref{section:federated_learning}). We consider full participation during FL rounds, meaning that all the users' updates are used for averaging in each communication step. Furthermore, we fix the batch size to $B=21$ (matching the size of local datasets) and the number of local epochs to $E=1$. Regarding data splits, we only use the scenario \textit{training3}, as it is the one containing more complex and complete data, and use the 80\% of the contexts for training, the 10\% of the contexts for the first validation, and the remaining 10\% for the second validation. Notice that we use the first validation set for early stopping of the global model, whereas the second one it to perform hyper-parameter tuning. We tune our method by using Tree Parzen Estimator of the Optuna library~\cite{akiba2019optuna} and choose the model with the lowest MAE on the second validation set.\ITUpar
	
	
	Finally, we evaluate our global model after every 20 rounds and stop training if no improvement in validation MAE is achieved after $T=100$ rounds. We evaluate the prediction results of our method via the MAE metric. Recall that we estimate the throughputs by multi-output regression task and each context may have different number of throughputs to be predicted. Thus, we flatten the predictions for existing STAs before calculating the MAE. We normalize the data by minimax normalization. We use the standard SGD implementation of Pytorch~\cite{paszke2017automatic} to implement federated averaging.\ITUpar
	
	\begin{table}[t!]
		\centering
		\caption{\textcolor{black}{Evaluation results for the best-performing neural networks with different numbers of layers in FederationS' hyperparameter optimization.}}
		\label{tab:fedipc_different_layers}
		\resizebox{.5\columnwidth}{!}{%
			\begin{tabular}{|c|c|c|}
				\hline
				\textbf{\# hidden layers} & \textbf{Neurons per layer} & \textbf{MAE (Mbps)} \\ \hline
				1 & 256 & 5.10 \\ \hline
				2 & 256, 16 & 5.57 \\ \hline
				3 & 256, 32, 16 & 5.88 \\ \hline
			\end{tabular}%
		}
	\end{table}
	
	\textcolor{black}{
		Table~\ref{tab:fedipc_different_layers} shows the evaluation results for neural network architectures with different number of layers. We only report the best configuration for each different number of hidden layers. We choose the hyperparameters with the least MAE on the first validation set and report the results on the second validation set for different numbers of layers. The network with only 1 hidden layer containing 256 neurons performs the best. As the number of layers increases, we observe a decrease in performance. This is probably because our network starts to overfit the data when the network becomes more complex. This reasoning also supports that, although our network is much simpler, it is more accurate than other participants' networks, as shown in Table~\ref{tab:summary_results}.}\ITUpar
	
	
	\subsection{WirelessAI}
	
	We follow the FL framework to address the complex SR problem in multiple 11ax WLAN cells. Individual agents first train their local NN models (with the same network structure) using their local datasets, and then exchange and average model weights through a centralized parameter server. \ITUpar
	
	Typical NN models use simple data structures such as vectors to encode inputs and outputs. However, in wireless networks characterized by graphs $G=(V,~E)$, where $V$ is the set of nodes, and $E$ is the set of wireless links, the number of nodes and the number of links can vary depending on the networking scenarios. It is difficult to fix the vector dimension to fit all networking scenarios. Even if we can fix the dimension and pad zeros to the unused dimension fields, it is meaningless to use these fields.\ITUpar
	
	To overcome the graph representation problem, we treat the whole network as an image. More specially, we first fix a maximum range and treat the whole network as a 1$\times$100$\times$100 gray-scale image with a default value of 0. Then we map nodes to values by their roles (i.e., AP role with value 1, the target AP with OBSS/PDD value, other APs with value 1, and STAs with value 2), and place the values to their corresponding locations. In this way, we can represent any networks with arbitrary APs and STAs. Note that the topology information is encoded into the image.\ITUpar
	
	Then, we adopt two NNs to predict the performance: one part is a convolutional neural network (CNN), and the other part is a fully connected neural network (FCNN). We first use CNN to capture the interactions between STAs and APs to predict the RSSI, SINR of the BSS of interest, and interference to the AP of interest. The input of the CNN is the above processed gray-scale image, and the used OBSS/PD value, and the output of the CNN is the RSSI, SINR, and the caused interference. Then, we use the output of CNN as the input of FCNN to predict the downlink throughput of the AP of interest. \ITUpar
	
	The key rationale of using such architecture is to reduce computation complexity. RSSI, SINR, and the caused interference are the key factors that impact the final performance. Compared with using a whole FCNN to predict the performance, if we can first use CNN to model the relationship among \{topology, OBSS/PD value\} and \{RSSI, SINR, and the caused interference\}, and then use a small dimension of FCNN to predict the performance, the computation complexity is reduced.\ITUpar
	
	To empower our FL algorithm, we treat each context as a local client and let each local client use its own data to train the above two NNs. In particular, we follow the standard FL training procedure, and run the training in rounds: the above two NN models are trained by each local client, and the weights of the local models are averaged to generate the global shared model, which is used in the next round. For a dataset, there are overall 1000 local clients, and we randomly choose 10 local clients in each round to generate the average model. The global shared model has been updated using $T=20$ rounds in total.\ITUpar
	
	The proposed NN model and FL algorithm are implemented in Pytorch.\footnote{The code used to implement all the proposed methods by WirelessAI is available in Github \cite{wirelessai2021repository}.} Table \ref{table_model_nn} and Table \ref{table_model_fc} summarize the architecture of the proposed NN models. In particular, there are 13 layers in our CNN model including 5 convolution layers, 4 max-pooling layers, 1 adaptive average pooling layer, and 3 fully-connected layers. For each convolution layer, the layers are convolved with kernel size 3. In order to keep the size of the image after each convolution operation and obtain more information on the image edge position, we fill the images (i.e., padding) before each convolution operation. After every convolution layer, a max-pooling operation is applied to the feature maps. The kernel size of the max-pooling layer is 2. The purpose of max-pooling is to reduce the size of the feature maps. The output size of the adaptive average pooling layer is 1. The fully-connected layers consist of respectively 512 and 64 and 17 output neurons. There are 1 input layer, 2 hidden layers, and 1 output layer in our FCNN model. These layers consist of respectively 512 and 128 and 64 and 6 output neurons. The ReLU is used as an activation function for convolution layers and fully-connected layers.\ITUpar
	
	\begin{table}[t!]
		\centering
		\caption{A Summary Table of the Proposed CNN Model.}
		\resizebox{.6\columnwidth}{!}{
			\begin{tabular}{|c|c|c|c|c|}
				\hline
				\textbf{Layers} & \textbf{Type} & \textbf{Output Size} & \textbf{Kernel Size} & \textbf{Stride} \\
				\hline
				1 & Convolution & 128$\times$100$\times$100 & 3 & 1 \\ \hline
				2 & Max-pooling & 128$\times$50$\times$50 & 2 & 2 \\ \hline
				3 & Convolution & 256$\times$50$\times$50 & 3 & 1 \\ \hline
				4 & Max-pooling & 256$\times$25$\times$25 & 2 & 2 \\ \hline
				5 & Convolution & 512$\times$25$\times$25 & 3 & 1 \\ \hline
				6 & Max-pooling & 512$\times$12$\times$12 & 2 & 2 \\ \hline
				7 & Convolution & 1024$\times$12$\times$12 & 3 & 1 \\ \hline
				8 & Max-pooling & 1024$\times$6$\times$6 & 2 & 2 \\ \hline
				9 & Convolution & 2048$\times$6$\times$6 & 3 & 1 \\ \hline
				10 & Adaptive average pooling & 1$\times$2048 & - & - \\ \hline
				11 & Fully-Connected & 512 & - & - \\ \hline
				12 & Fully-Connected & 64 & - & - \\ \hline
				13 & Fully-Connected & 17 & - & - \\ \hline
		\end{tabular}}
		\label{table_model_nn}
	\end{table}
	
	\begin{table}[t!]
		\centering
		\caption{Summary Table of the WirelessAI FCNN Model.}
		\begin{tabular}{|c|c|c|}
			\hline
			\textbf{Layers} & \textbf{Type} & \textbf{Output Size} \\
			\hline
			1 & Fully-Connected & 512 \\ \hline
			2 & Fully-Connected & 128 \\ \hline
			3 & Fully-Connected & 64 \\ \hline
			4 & Fully Connected & 6 \\ \hline
		\end{tabular}
		\label{table_model_fc}
	\end{table}
	
	\section{Performance Evaluation}
	\label{section:performance}
	
	In this section, we show the results obtained by the challenge participants' models presented in Section~\ref{section:solutions}. During the competition, the test dataset was released without revealing the actual throughput of the simulated deployments. Table~\ref{tab:summary_results} summarizes the performance accuracy obtained by each participating team on the test dataset. \ITUpar
	
	\begin{table}[t!]
		\centering
		\caption{Mean average error obtained by the solution proposed by each team.}
		\label{tab:summary_results}
		\begin{tabular}{|c|c|}
			\hline
			\textbf{Team} & \textbf{MAE (Mbps)} \\ \hline
			FederationS & 6.5534 \\ \hline
			FedIPC & 5.8572 \\ \hline
			WirelessAI & 8.913 \\ \hline	
		\end{tabular}
	\end{table}
	
\begin{figure}[tb!]
		\centering
		\includegraphics[width=.45\columnwidth]{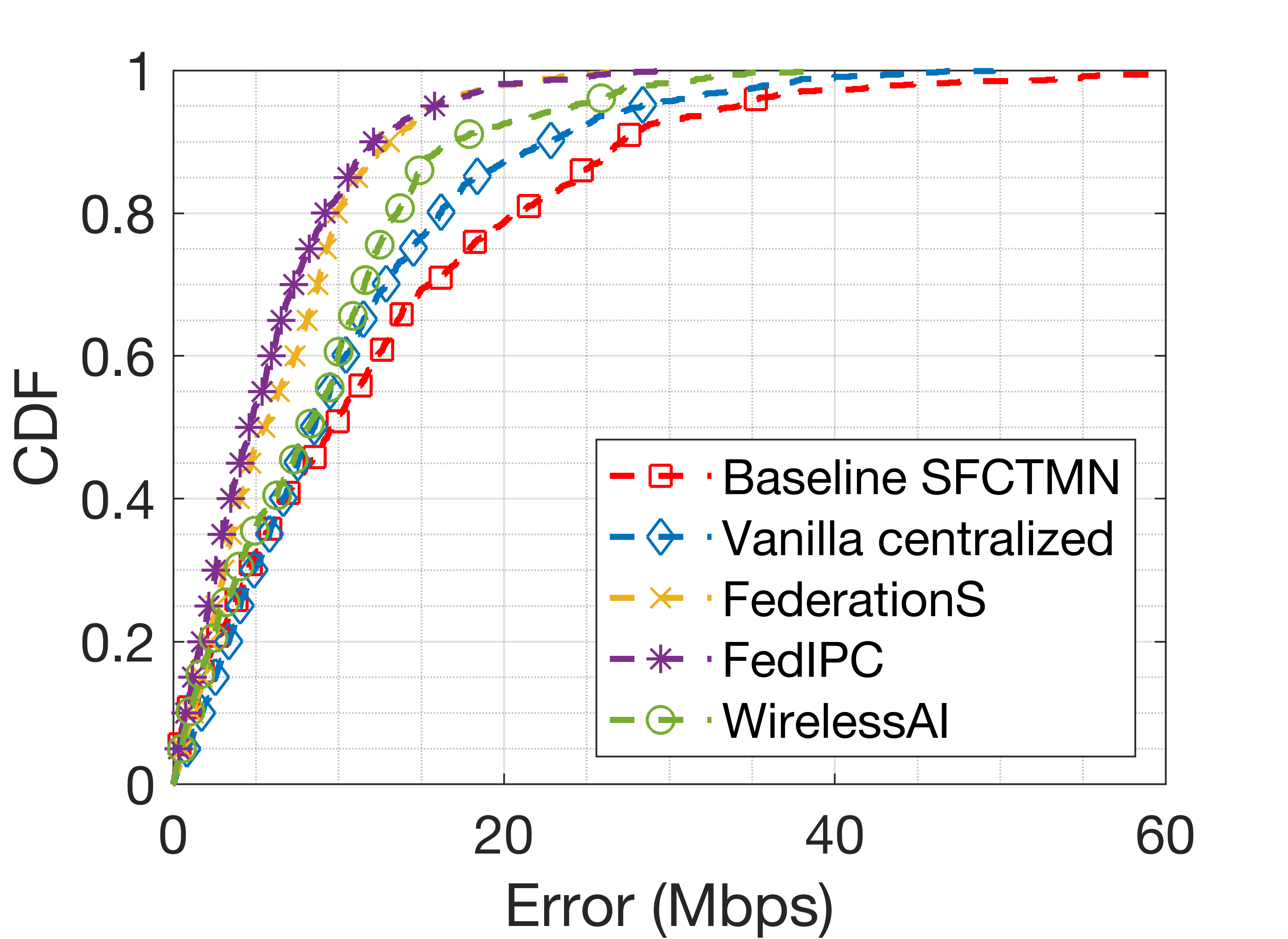}
		\caption{Empirical CDF of the error obtained by each participants' solution over the test dataset. \textcolor{black}{Results of baseline SFTCMN modeling and vanilla centralized NN are included for comparison purposes.}}
		\label{fig:summary_results}
	\end{figure}

	Next, we analyze the results obtained by each participant in more detail. First, Fig.~\ref{fig:summary_results} showcases the empirical cumulative distribution function (CDF) of the test error obtained by each solution. The results are compared to the ones obtained by a vanilla centralized mechanism, which consists of a feed-forward NN with $1024$, $512$, and $256$ neurons in each of its three layers, with ReLU activation.\footnote{For further details on the centralized mechanism, refer to the provided open-access repository~\cite{github_challenge}.} \textcolor{black}{In addition, to remark the need for ML for the prediction problem in Wi-Fi, we provide the results of a baseline analytical model, based on Continuous Time Markov Networks (CTMNs). Such a baseline model implementation was presented in~\cite{wilhelmi2021spatial}, as an extension of the Spatial Flexible Continuous Time Markov Network (SFCTMN) framework~\cite{barrachina2019sfctmn}. To the best of our knowledge, the targeted analytical implementation of IEEE 802.11ax SR is one of the first of its kind and suits the SR operation because it characterized both PHY and MAC phenomena in BSSs.}\ITUpar
	
	As shown, all the proposed FL solutions improve the performance \textcolor{black}{of both the SFCTMN analytical model and the naive centralized method. The SFCTMN model, while allowing to provide insights on the 11ax SR operation, is shown to fail at faithfully representing the realistic phenomena observed in WLANs through simulations. This result points to the need for ML models to capture the complex phenomena in WLANs.} Concerning the ML models presented in this paper, FedIPC is the one providing the highest accuracy, with the 82.4\% of the predictions below a 10 Mbps error, compared to the 80.5\% and the 60.3\% achieved by FederationS and WirelessAI, respectively.\ITUpar
	
\begin{figure}[tb!]
		\begin{subfigure}[b]{.5\linewidth}
			\centering
			\includegraphics[width=\textwidth]{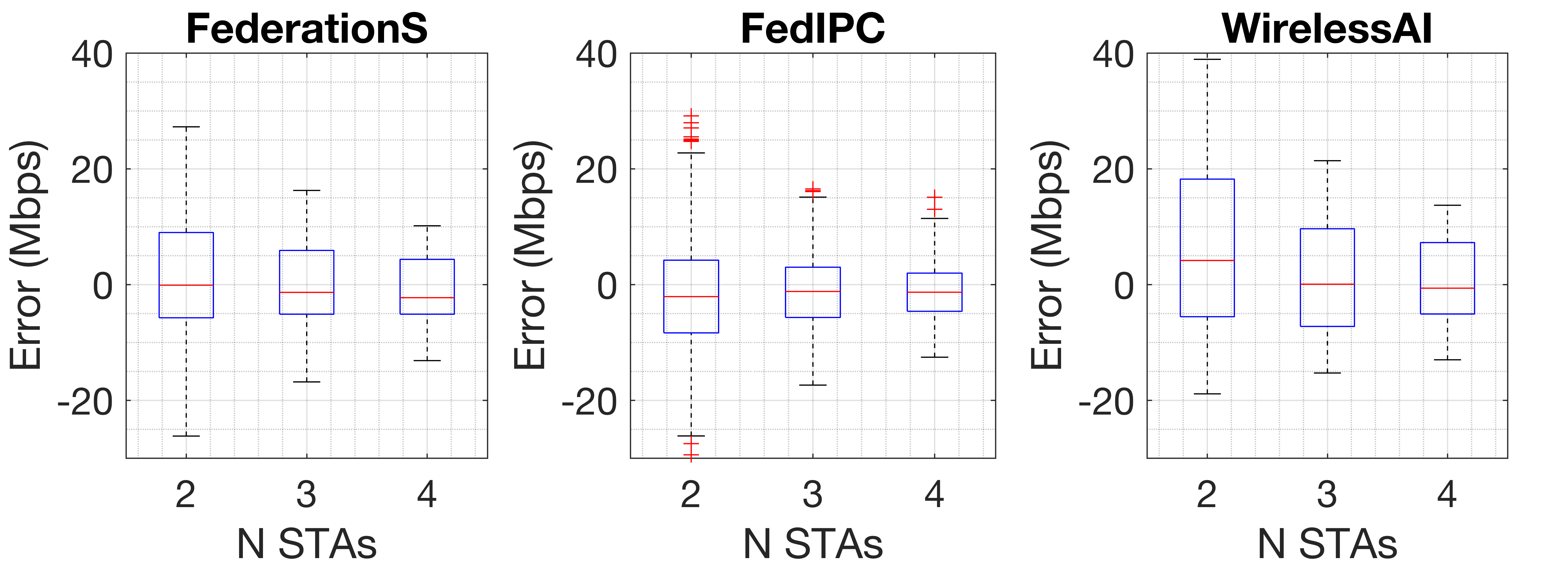}
			\caption{}
			\label{fig:boxplot_num_stas}
		\end{subfigure}
		\begin{subfigure}[b]{.5\linewidth}
			\centering
			\includegraphics[width=\textwidth]{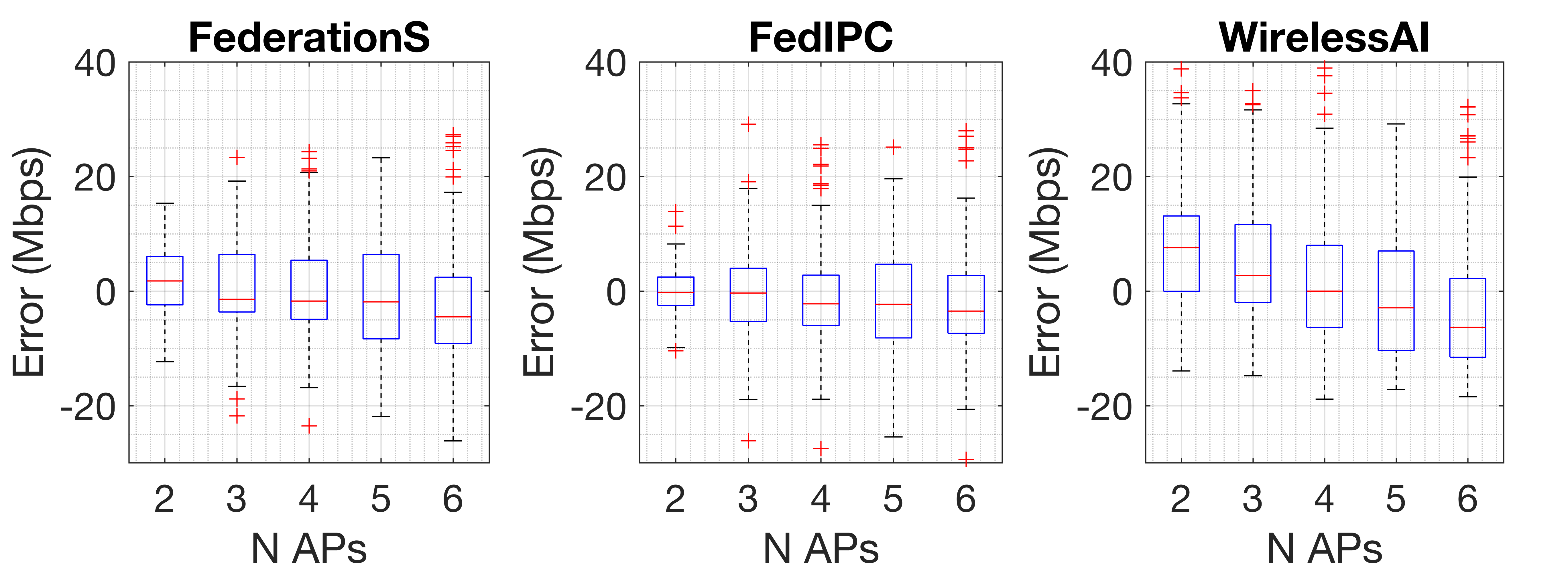}
			\caption{}
			\label{fig:boxplot_num_aps}
		\end{subfigure}
		\caption{Generalization capabilities of the proposed models on test data with respect to: (a) the number of STAs, and (b) the number of APs.}\label{fig:boxplot_complexity} 
	\end{figure}

	Finally, to provide more insights on the generalization capabilities of each model, \textcolor{black}{Fig.~\ref{fig:boxplot_complexity}} shows the error obtained by each solution, for each possible number of APs and STAs in the test deployments. As shown, contrary to intuition, all the models perform better as complexity increases, i.e., as the number of APs and STAs is bigger. This result is mainly motivated by the fact that DL allows capturing the most complex interactions in dense deployments, which reinforces the role of AI-enabled solutions for network optimization.
	
	\section{Discussion}
	\label{section:conclusions}
	
	\subsection{Contributions}
	
	In this paper, we have presented the main results gathered from problem statement \textit{``ITU-ML5G-PS-004: Federated Learning for Spatial Reuse in a multi-BSS (Basic Service Set) scenario''} in the 2021 ITU AI for 5G Challenge. First, we have overviewed the SR problem in IEEE 802.11ax and formulated a novel optimization use case via FL. To evaluate the potential of this solution, we have provided a dataset containing synthetic data on 11ax SR measurements in random deployments. The dataset is open for the sake of reproducibility and to engage other researchers to work on this topic. The provided dataset has been used by the participants of the challenge to develop the models introduced in this paper.\ITUpar 
	
	
	\subsection{Lessons learned}
	
	We extract the following insights from the models and results overviewed in this paper:
	\begin{enumerate}
		\item Predicting WLAN performance accurately is key to optimize these kinds of networks. However, mechanisms like OFDMA or SR add further complexity to accurately modeling WLANs. In this regard, DL-based models have shown a great potential for capturing the complex interactions of IEEE 802.11ax WLANs applying SR in dense deployments. This provides a paradigm shift with respect to mostly adopted online learning mechanisms. Nevertheless, for the sake of addressing spatial interactions in dynamic WLAN settings, both types of mechanisms are envisioned to be combined.
		\item FL suits the decentralized nature of WLANs and, despite contexts count on limited data, its performance has been shown to outperform vanilla centralized ML. Thus, FL provides opportunities for (i) training ML models collaboratively, (ii) enhancing user privacy by not sharing data directly but model weights, (iii) reducing the communication overhead of traditional centralized ML mechanisms, and (iv) providing portability by reducing the computation capabilities for the training of ML models. 
		\item Finally, using synthetic datasets for training ML models contributes to enriching ground knowledge on certain network technologies and deployments. Concerning this, we remark the importance of cost-effectively generating data with network simulators, which can complement real networking data to, for instance, reproduce anomalies.
	\end{enumerate}
	
	


	\subsection{Future research directions}
	
	The SR mechanism is expected to evolve towards a more sophisticated operation in future IEEE 802.11 amendments. At the moment of writing this paper, task group IEEE 802.11be (TGbe) is defining the coordinated SR (c-SR) mechanism as part of the multi-AP operation~\cite{nunez2021txop}. Through c-SR, APs collaborate to further improve the performance gains achieved by applying SR. More specifically, APs exchange relevant information (e.g., measurements) to select the best SR configuration, based on the recipient STAs to which transmissions are expected to be held.\ITUpar
	
\begin{figure}[tb!]
		\begin{subfigure}[b]{\linewidth}
			\centering
			\includegraphics[width=.35\textwidth]{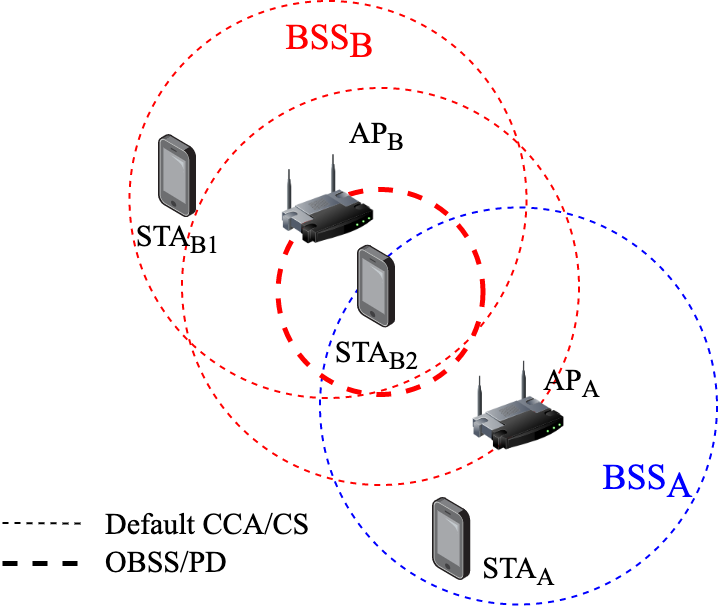}
			\caption{}
			\label{fig:csr_1}
		\end{subfigure}
		\begin{subfigure}[b]{\linewidth}
			\centering
			\includegraphics[width=.45\textwidth]{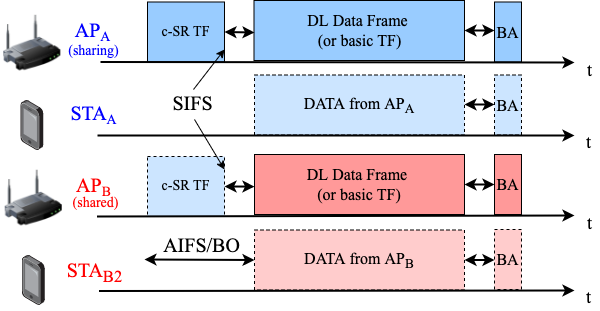}
			\caption{}
			\label{fig:csr_2}
		\end{subfigure}
		\caption{c-SR operation: (a) deployment, and (b) exchange of packets.}\label{fig:csr} 
	\end{figure}

	Fig.~\ref{fig:csr} illustrates the basics on c-SR. Considering the deployment shown in Fig.~\ref{fig:csr_1}, where two contending APs (namely AP$_A$ and AP$_B$) are within the same sensitivity area when using the default CCA/CS. Nevertheless, when applying c-SR, both APs can transmit in parallel, thus enhancing spectral efficiency and potentially reducing latency. To do so, the AP gaining channel access after completing the backoff (BO) takes the role of \textit{sharing AP} (in Fig.~\ref{fig:csr_2}, AP$_A$ is the sharing AP). Likewise, AP$_B$ acts as a \textit{shared AP}. The sharing AP sends a c-SR Trigger Frame (TF) to indicate the capabilities of the upcoming transmission to selected STA (STA$_A$), including the maximum acceptable interference level (obtained through measurements). Based on the information provided in the TF, the shared AP decides which STA to transmit and selects the best configuration (e.g., transmit power, MCS) to that end. Notice, as well, that a transmit power limitation is imposed by the sharing AP so that its transmission is not affected by shared APs. Finally, once both simultaneous transmissions take place, a block ACK (BA) is sent by STAs to confirm successful downlink transmissions.\ITUpar
	
	Given the added complexity of evolved SR, the role of ML, and more specifically FL, gains ground. Notice, as well, that in order to set the best configuration that maximizes the overall network performance, both APs and STAs perform measurements related to spectral utilization. Such a rich source of data can be exploited by FL to drive intelligent-based network optimization.
	
	
	\section*{Acknowledgement}
	\label{sec:ackn}
	The authors would like to thank enormously everyone that made possible the ITU AI for 5G Challenge, with special mention to Vishnu Ram OV, Reinhard Scholl, and Thomas Basikolo. Likewise, we would like to thank the valuable feedback provided by Dr. Andrea Bonfante.
	
	The present work has received funding from the European Union’s Horizon 2020 Marie Skłodowska Curie Innovative Training Network Greenedge (GA. No. 953775) and has been partially supported by WINDMAL PGC2018-099959-B-I00 (MCIU/AEI/FEDER,UE). This work was also in part funded by the SFI-NSFC Partnership Programme Grant Number 17/NSFC/5224.

\end{document}